\documentclass[12pt,nofootinbib,epsfig,superscriptaddress,showpacs]{article}
\usepackage{cite} 
\usepackage[utf8]{inputenc}

\usepackage{graphicx}
\usepackage{amsmath}
\usepackage{amssymb}
\usepackage{cmap}
\usepackage{bm}

\usepackage[left=2.5cm,right=2.5cm,top=2.5cm,bottom=2.5cm]{geometry}

\newcommand{\be}{\begin{equation}}
\newcommand{\ee}{\end{equation}}

\newcommand{\ben}{\begin{equation*}}
\newcommand{\een}{\end{equation*}}

\newcommand{\bea}{\begin{eqnarray}}
\newcommand{\eea}{\end{eqnarray}}

\def\la{\mathrel{\mathpalette\fun <}}

\def\fun#1#2{\lower3.6pt\vbox{\baselineskip0pt\lineskip.9pt
\ialign{$\mathsurround=0pt#1\hfil ##\hfil$\crcr#2\crcr\sim\crcr}}}

\newcommand{{\SD}}{\rm SD}

\newcommand{\vex}{\mbox{\boldmath${\rm x}$}}
\newcommand{\vey}{\mbox{\boldmath${\rm y}$}}

\newcommand{{\Mc}}{\mathcal{M}}

\newcommand{\lan}{\langle}
\newcommand{\ran}{\rangle}

\begin{document}

\title{\bf Dynamical role of Polyakov loops in the QCD thermodynamics}

\author{
N.O.~Agasian$^{a,b,}$\thanks{agasian@itep.ru}\,\,, M.S.~Lukashov$^{a,c,}$\thanks{lukashov@phystech.edu}\,\, and Yu.A.~Simonov$^{a,}$\thanks{simonov@itep.ru}\, \\
\\ 
$^a$ \small{\em Alikhanov Institute for Theoretical and Experimental Physics,}\\
\small{\em Moscow 117218, Russia}\\
$^b$ \small{\em National Research Nuclear University ``MEPhI'',}\\
\small{\em Moscow 115409, Russia}\\
$^c$ \small{\em Moscow Institute of Physics and Technology,}\\
\small{\em Dolgoprudny 141700, Moscow Region, Russia}
}\bigskip

\date{\today}

\maketitle

\begin{abstract}
Polyakov loops  $L_a(T), a=3,8,...$ are shown to give the most  important
nonperturbative contribution to the thermodynamic  potentials. Derived from the
gluonic  field correlators they enter as factors into free energy. It is shown
in the  $SU(3)$ case  that $L_a (T)$ define to a large extent the behavior of the
free energy and the trace anomaly $I(T)$, most sensitive to
nonperturbative effects. \smallskip


\rule{0mm}{5mm}

 \medskip

 \bigskip
\end{abstract}

\newpage

{\bf 1.} Polyakov lines (PL) $L_a(T), a=3,8,\ldots$ play a double role in the dynamics of
hot QCD. First of all, they serve as an order parameter (see \cite{1a,1b} for
reviews) being nonzero above the critical temperature and signalling the
absence of confinement (e.g. for the  adjoint
 PL  in the SU(3) theory, there is a strong jump in  the  values of PL at $T=T_c $ \cite{2}).
 Secondly, as we stress below, PL have an important role in the whole dynamics of the hot QCD. In
 the Field  Correlator (FC)  approach this was directly derived from the basic
 QCD Lagrangian with account of the quadratic gluon field correlators \cite{3, 4}. It was shown in \cite{3,4} that the free energy is proportional to
 the  $L^n$ in the  Matsubaru series over $n$.  As will be shown below, this
 dependence is  crucial in defining behavior of all thermodynamic quantities in
 the region $T_c \leq T \la 4T_c$  and is substantial for $T\la 10 T_c$. In
 particular, the  remarkable plateau  of $\frac{I(T)}{T^2T^2_c}$ in SU(3),
 discovered in \cite{5}, is for  the most part due to the $1/T^2$ behavior of
 $T\frac{\partial}{\partial T} L_{adj} (T)$.

 In other approaches to the hot QCD dynamics the role of PL was also taken
 into account in different ways, e.g. in the matrix PL models \cite{6}, and
 in the PNJL model \cite{7a,7b,7c}, introducing  an additional potential $V(L,L^+$) in
 the  Lagrangian, see \cite{8} for a review. \bigskip

{\bf 2.}  The quadratic gluon field  correlator  consists of two  colorelectric terms
  $D^E$ and $D^E_1$ \cite{9},
\begin{gather}
D_{\mu\nu \lambda \sigma}(x,y) \equiv  g^2 tr_a\lan F_{\mu} (x) \Phi
  F_{\lambda\sigma} (y) \Phi\ran =\notag\\
= c_a \left\{\vphantom{\frac12}(\delta_{\mu\sigma} \delta_{\nu\sigma}-
  \delta_{\mu\sigma} \delta_{\nu\lambda}) D (x-y) + \right. \notag\\
\left. +\frac12 \left[ \vphantom{\frac12}
  \frac{\partial}{\partial x_\mu}(x_\lambda\delta_{\nu\sigma} - x_\sigma \delta_{\nu\lambda})
  + (\mu\lambda\leftrightarrow \nu\sigma)\vphantom{\frac12} \right] D_1 (x-y)\vphantom{\frac12} \right\},
\label{1}
  \end{gather}
  and  the resulting  nonperturbative (np) plus perturbative  interaction between color objects in the repr. $a$ can
  be written as \cite{9}
\begin{gather}
  V_a (r) =c_a \left\{ 2\int^r_0 (r-\lambda) d\lambda \int^\infty_0 d\nu D^E
  (\lambda,\nu) +\right.\notag\\
\left.+ \int^r_0 \lambda d\lambda \int^\infty_0 d\nu D_1^E(\lambda,\nu)\right\} = \notag\\ = c_a \left\{ V_{\rm conf} (r) + V_1 (r) \right\},
  \,c_3 =1,\,c_8=\frac{9}{4},~\text{\rm  etc.}
\label{2}
\end{gather}

  In the deconfinement phase $(V_{\rm conf} =0)$, $V_1(r)$ has an important
  property that $V_1(\infty) =$const, which implies, that each deconfined gluon
  (or  quark) goes astray with a piece of energy  $\frac{c_a}{2} V_1 (\infty)$. It is important, that
  this term appears in  the gluon pressure in the exponent,    $\exp \left( -c_a \frac{V_1(\infty)}{2T}\right)$
  as follows from
  the path integral form of the gluon pressure \cite{3}
\be P_{gl} = (N^2_c -1) \int^\infty_0 \frac{ds}{s} \sum_n G^{(n)} (s),
\label{3}\ee where $G^{(n)}(s)$ is the winding path integral over the loop
$C_n$, where all gauge field dependence  enters as \begin{gather} G^{(n)} (s) \sim  \lan
\exp (ig \int_{C_n}\!\!\! dz_\mu A_\mu)\ran = \exp \left( -\frac12 \int_{S_n}\!\!\! d
\sigma_{\mu\nu}(u)\times \right. \notag \\ \left. \times\int_{S_n}  d\sigma_{\lambda\sigma} (u) \lan F_{\mu\nu} \Phi
F_{\lambda\sigma} \Phi\ran+O(F^4)\right).\label{4}\end{gather}

Insertion of the field correlator (\ref{1}) in (\ref{4})  produces exactly the integral \ben
J (T, r ) = \exp \left( - \frac{c_a}{2} \int^{1/T}_0  dt_E V_1 (r, T) \right) =\een \be=
\exp \left( - \frac{c_a}{2T}  V_1 (r, T) \right).\label{5}\ee

 Following \cite{10}, it is convenient to extract from $V_1 (r,T)$ the large
 distance limit $V_1 (\infty, T)$, leaving the sum of the  attractive
 interactions $\Delta V_1 = V_1 (r, T) - V_1(\infty, T)$ and the renormalized
 perturbative interaction $V_1^C (r,T)$ to account for  as a correction. As a
 result in the leading approximation the function $J(T, r)$ in  (\ref{5})
 acquires a factor $J(T, \infty)$, entering
 in $G^{(n)}(s)$ and $P_{gl} (T)$, which we call the Polyakov
line $L_a (T)$\be L_a (T) = \exp \left( - \frac{c_a}{2} \frac{V_1(\infty,
T)}{T}\right)\label{6}\ee \be G^{(n)}(s) = \frac{1}{ \sqrt{4\pi s}} e^{-\frac{n^2}{4sT^2}} G_3 (s)
L_8^{n} (T), \label{7}\ee and $G_3(s)$ is the 3d path integral over the 3d
 portion of the loop $C_n$ \be G_3(s) = \int (D^3z)_{xx} e^{-K_{3d}} \lan W_3
\ran \label{8}\ee

Here the 3d projected Wilson loop $\lan W_3\ran$ obeys the spatial area law
with the colormagnetic string tension  $\sigma_s$ and  the 3d  area $A_3$ \be
\lan W_3\ran =\exp (-\sigma_s A_3).\label{9}\ee

In \cite{3,4} $G_3 (s)$ was calculated in the approximation when $\sigma_s =0$,
and as a result one has $G_3 (s) =\frac{1}{ (4\pi s)^{3/2}}$, and
\bea
P^{(0)}_{gl} &=& \frac{2(N^2_c -1)}{\pi^2} T^4 \sum^\infty_{n=1} \frac{1}{n^4}
L^n_8 = \nonumber \\ &=& \frac{2(N^2_c -1)}{\pi^2} T^4 {\rm\ Li}_4(L_8)
\label{10}
\eea which for $L_8=1$ yields the Stefan-Boltzmann result
$P_{gl}^{(SB)}  = \frac{(N^2_c -1) \pi T^4}{45}$,  defining the asymptotic
behavior of $P_{gl}$, when $V_1$ decreases at large $T$.

Note several important points in our definition of $L_a (T)$:

i) $L_a (T)$ automatically satisfies the Casimir scaling law due
to factor $c_a$ in (\ref{2}), this scaling is supported by lattice
data \cite{2,10x}.

ii) In the correlator $P(\vex-\vey)$ of two Polyakov loops,   studied in
\cite{10}, one obtains the same form as in (\ref{4}) with the loops $(S_n,
S_n)\to (S_n, S'_n)$ referring to  two different loops at the distance $r =
|\vex-\vey|$ from each other, and one obtains the same form as in \cite{11a,11b}\bea
P (\vex-\vey) &=& \frac{1}{N_c^2} \exp \left( - \frac{\tilde F_1(r, T)}{T}
\right) + \nonumber \\ &+& \frac{N^2_c-1}{N^2_c} \exp \left( - \frac{\tilde F_8 (r, T)}{T}
\right), \label{11}\eea where e.g. $\tilde F_1 (r,T) = c_a(V_1 (r,T)+ V_{conf}
(r,T))$. As a result $P(r )$ vanishes in the  confining phase for $r  \to
\infty$ and is a product of two Polyakov loops in this limit in the deconfined
phase, as it should be. This exercise also implies that the Polyakov loop
enters in $P_{gl}^{(0)}$, Eq. (\ref{10}), in the approximation, when the
interaction $V_1(r , T)$ between neighboring gluons is replaced by $V_1(\infty,
T)$.

iii) The definition  (\ref{6}) of PL  appears due to the vacuum average of
gluonic field, Eq. (\ref{1}), which evidently violates the $Z(3)$ symmetry.\bigskip

{\bf 3.} As  was  stated above the resulting gluon pressure $P_{gl}$ in the lowest
approximation   is given by \bea P_{gl} &=& \frac{(N^2_c-1)}{\sqrt{4\pi}}
\!\int^\infty_0 \!\!\!\frac{ds}{s^{3/2}}\times \nonumber \\ & \times &\!\!\!\sum_{n=1,2,\ldots}\!\!\!e^{-\frac{n^2}{4s T^2}}\,G_3
(s) L_8^{(n)} (T),\label{12}\eea and the point is where to find the information
about PL. This  can be obtained from several sources:

a) From the lattice data on $D_1(x)$ in \cite{12a,12b,12c}, where it was found that the
correlator $D_1(x)$, unlike $D(x)$, does not vanish above $T_c$, and decays as
$\exp (-M|x|)$  with $M=O(1$ GeV).

The corresponding values of $V_1 (R,T)$ were calculated from $D_1$ in the
interval  $1.007 \leq T/T_c\leq 1.261$ in \cite{13}, however with low accuracy.

b) From the gluelump representation of $D_1 (x)$ in \cite{14} one finds in
\cite{10} that the np part of $V_1$ can be represented as  \begin{gather} V_1^{(np)}(\infty,T) = \frac{d}{M_1} \left[1-\frac{T}{M_1}
\left(1-e^{-M_1/T}\right)\right], \label{13} \\ d=0.432~{\rm GeV}^2, ~ M_1=0.69~{\rm
GeV}. \notag \end{gather}

This form agrees with lattice data \cite{15} and   can be used to define
$L_a(T)$  at least for $T<2T_c$.

c) From the free energies $F_i (r, T)$,  obtained from the PL correlator
\cite{15,11a,11b}, which have the same form as in (\ref{11}) with the  replacement
$\tilde F_i\to F_i$. This replacement implies, as was stated in \cite{10}, that
the lattice version of $V_1 (r,T)$ is the singlet free energy $F_1(r,T)$, which
is an averaged value over all excited states,  yielding the inequality
$F_1(r,T) < V_1 (r,T)$. As a result one obtains $L_a(T)$ in (\ref{6}), which satisfies the condition
$L_a(T)<L^{lat}_a(T)$, where $L^{lat}_a(T)$ is found on the lattice via $F_1(\infty,T)$.
In particular, $F_1(\infty,T)$ becomes negative for $T>2T_c$, yielding $L^{lat}_a(T)>1$,
while in our case for all $T$ $L_a(T)<1$. In what follows we are using the form $V_1(T)$,
which is close to that in \cite{3,4} and the resulting $L_a(T)$ is close the lattice data of \cite{2} for $T\leq2T_c$, namely

\be
V_1(\infty, T) =\frac{0.13~{\rm GeV}}{ T/T_c-0.84}.
\label{14}
\ee

\smallskip

{\bf 4.} In the previous  section we have disregarded the colormagnetic interaction (CM)
 contained in $G_3(s)$ in (\ref{12}).  To account for the CM effects, one
 should calculate $G_3 (s)$ in (\ref{8}), where $K_{3d} = \frac14 \int^s_0
 \sum^3_{i=1} \left( \frac{dz_i}{d\tau}\right)^2 d \tau$. As it is seen from
 (\ref{8}), what one  should estimate is the gluon loop in 3d, covered with the
 confining film with string tension $\sigma_s (T)$. Using the same method as in \cite{17a,17b},
  one can calculate $G_3(s)$ in terms of the 2d gluon-gluon bound
 states with masses $M_\nu = 4 \omega_\nu^{(0)}$, where $\omega_\nu^{(0)}
 =\frac32 \left( \frac{a_\nu}{3}\right)^{3/4} \sqrt{\sigma(T)},
 \nu=0,1,2,...,  a_0 =1.74$ namely
 \be G_3(s) =\frac{1}{\sqrt{\pi s}} \sum_{\nu=0,1,..} \varphi^2_\nu (0)
 e^{-M_\nu\omega_\nu^{(0)} s},\label{15}\ee
 where $\varphi_\nu(0)$ is the 2d wave function at origin. From dynamical
 consideration $\varphi^2_\nu (0) = c_\nu\sigma_s (T)$, with $c_\nu$ --
 numerical constant. Moreover, $ M_\nu \omega_\nu^{(0)} \cong 4 \sigma_s (T)
 \approx m^2_D (T)$, where $m_D(T)$ is the np Debye screening mass, calculated
 in \cite{17a,17b} in agreement  with lattice data \cite{18}. Thus keeping the
 lowest term with $\nu=0$ in (\ref{15}) one has $G_3^{(min)} (s) =
 \frac{1}{\sqrt{\pi s}} c_0 \sigma_s e^{-m^2_Ds}$ and inserting this into
 (\ref{12}) one has

 \bea P^{(min)}_{gl} (T)&=&\frac{(N_c^2-1) c_0 \sigma_s m_DT}{2\pi^2}\times \nonumber \\ &\times&\!\!\!\sum_{n=1,2,\ldots}\!\!\!\frac{1}{n} K_1 \left( \frac{n m_D}{T}\right)
 L_8^n.\label{16}\eea
It was shown in \cite{19a,19b} that
\be
\sqrt{\sigma_s(T)}= c_\sigma g^2 (T) T,\label{21a}
\ee
where use was made of the two-loop expression for $g^2(T)$
\begin{gather}
g^{-2}(T) = 2 b_0 \ln \frac{T}{\Lambda_\sigma}+
\frac{b_1}{b_0}\ln \left(2\ln \frac{T}{\Lambda_\sigma}\right), \label{22a} \\
b_0=\frac{11N_c}{48\pi^2},~~~ b_1=\frac{34}{3} \left(\frac{N_c}{16\pi^2}\right)^2. \notag
\end{gather}
The two constants $c_\sigma$ and $\Lambda_\sigma$ were determined
using a two-parameter fit to lattice results. For the  SU(3) gauge theory
$c_\sigma=0.566\pm 0.013,~~ \Lambda_\sigma=(0.104\pm 0.009) T_c$
\cite{19a,19b}.

 At large $T, \sigma_s(T)$ behaves as $c^2_\sigma g^4 (T) T^2$ , where $g^2(T)$
 is $O\left( \dfrac{1}{ln \frac{T}{\Lambda_\sigma}}\right)$(however $c_\sigma$ is a np quantity \cite{20}), and as a result
 $P_{gl}^{(min)} (T) / T^4 \sim O\left(
 \dfrac{1}{ln^2\frac{T}{\Lambda_\sigma}}\right)$.This  amounts to the approximately
 50\% decrease of $P^{(min)}_{gl} $ from $T=2T_c$ to $T=5T_c$, therefore it is important to
 consider also the higher states in the  sum over $\nu$.

 To  account for higher states it is  convenient to exploit the oscillator form
 of the  colormagnetic interaction, which immediately produces the analytic
 answer, namely

 \be G_3  (s) =  \frac{1}{(4\pi s)^{3/2}} \frac{M_0^2}{{\rm sh} M_0^2 s}, \label{17}\ee
where $M_0=\omega$ in the lowest excitation in the oscillator potential, which
we can associate with the screening mass $m_D = 2 \sqrt{\sigma_s}$ \cite{17a,17b}.

Inserting (\ref{17}) in (\ref{12}) one obtains the final form of the gluon
pressure with account of the spatial confinement in the oscillator form
 \be P^{(osc)}_{gl}  = \frac{2(N^2_c-1)}{(4\pi)^2}
 \sum_{n=1 }^\infty  L^n_8 \int \frac{ds}{s^2} e^{-\frac{n^2}{4sT^2}}
\frac{M_0^2}{{\rm sh} M_0^2 s}. \label{18}\ee Note, that in the limit $M^2_0 \to 0$ one recovers the free case, Eq. (\ref{10}). 

One can also use the oscillator form, reproducing the linear confinement with
the accuracy of 5\%; this  corresponds to the  replacement in
(\ref{18}): $ \frac{M^2_0}{{\rm sh} M^2_0 s} \to \frac{1}{s} \left( \frac{M^2_0 s
}{{\rm sh} M^2_0 s} \right)^{1/2}$. This modified oscillator
form we are using below in our calculations. However the final result is almost (within few
percent) insensitive to this replacement. \bigskip


{\bf 5.} The results of numerical calculation of the pressure in the approximations:
$ P^{(0)}_{gl} (T)$ and $P^{(osc)}_{gl} (T)$  with $L_8 (T)$ using
(\ref{14}) are given in the Fig. 1, in comparison with the   lattice data of
\cite{5}. One can see an improvement of the results, when $\sigma_s(T)$ is
taken into account in $P_{gl}^{(osc)} (T)$, however already $P_{gl}^{(0)} (T)$,
where only $L_8(T)$ is taken into account, is a reasonable approximation. This
supports our main idea, that PL are the important dynamical input, which should
enter $P_{gl}$ as factors, according to our derivation.

{
\begin{figure}[htb] 
\setlength{\unitlength}{1.0cm}
\centering
\begin{picture}(8.0,4.1)
\put(0.5,0.5){\includegraphics[height=3.5cm]{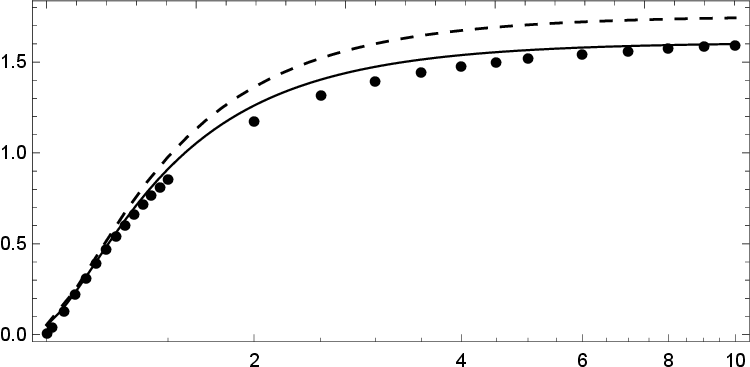}}
\put(7.05,0.1){$T/T_c$}
\put(0.1,3.25){\rotatebox{90}{$P/T^4$}}
\end{picture}
\caption{The pressure $\frac{P(T)}{T^4}$ in the SU(3) theory.
The dashed line corresponds to the pressure without magnetic confinement
Eq.(\ref{10}). The solid line is for the modified oscillator confinement, and filled dots are for the lattice data \cite{5}.}
\label{fig:fig01}
\end{figure}
}\medskip

Leaving details of comparison, as well as entropy $s(T)$, internal energy
$\varepsilon (T)$ and sound velocity $c_s(T)$ to another publication \cite{21},
we shall consider in more detail the scale anomaly $I(T) = \varepsilon -3 P$,
which can be written as
\be \frac{I(T)}{T^4} = T\frac{\partial}{\partial T} \left( \frac{P_{gl}}{T^4}
\right) = \frac{\bar I(T)}{T^4} +   p_{gl}(T) \frac{T\partial L_8}{\partial
T}\label{20}\ee
where we write $P_{gl}$ as $\frac{P_{gl}}{T^4} = p_{gl} L_8
(T)$, and $$ \bar I(T) = T\frac{\partial p_{gl}}{\partial T} L_8 (T).$$

In Fig. 2  we show $\frac{ I(T)}{T^4}$ and $\frac{I(T)}{T^4} (\frac{T}{T_c})^2$ as
functions of $T$ in the interval $T_c\leq T \leq 10 T_c$, and  note, that  as
was found on the lattice in \cite{5}, this purely np phenomenon, discovered in
\cite{5} is well reproduced by mostly the properties of $\frac{\partial L_8
(T)}{\partial T}$ which behaves in this region as $1/T^2$.

{
\begin{figure}[htb] 
\setlength{\unitlength}{1.0cm}
\centering
\begin{picture}(8.4,4.8)
\put(0.55,0.425){\includegraphics[height=4.5cm]{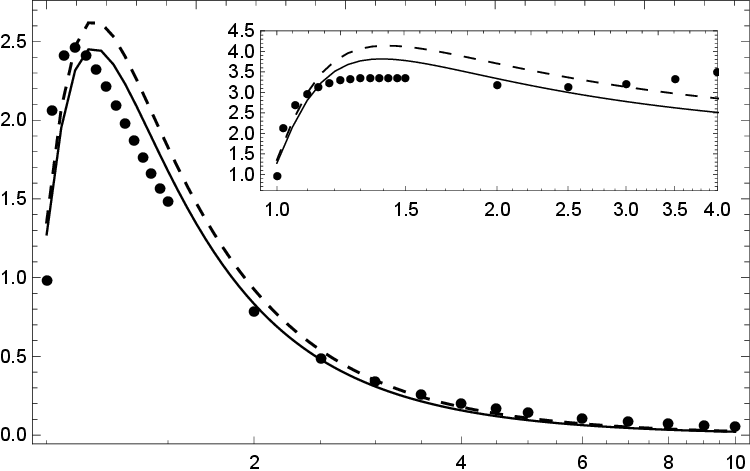}}
\put(7.1,0.1){$T/T_c$}
\put(0.2,4.1){\rotatebox{90}{$I/T^4$}}
\end{picture}
\caption{\small The trace anomaly $\frac{I(T)}{T^4}$. Notations are the same as in Fig.1.
In the upper right corner the plot is given for $\frac{I(T)}{T^4} (\frac{T}{T_c})^2$.}
\label{fig:fig02}
\end{figure}
}\medskip

Our purpose in this paper was to demonstrate the dynamical importance of the
Polyakov loops in the QCD thermodynamics in the SU(3) case. We have also shown
in some detail that PL enter thermodynamic potentials as factors and contain a
most part of np dynamics, which allows to explain the spectacular shoulder in
the $\frac{I(T)}{T^4} (\frac{T}{T_c})^2$ dependence.\smallskip

This work was  supported  by a grant from the Russian Science Foundation
(project number 16-12-10414).

\end{document}